\documentclass[aip,jcp,reprint,showkeys]{revtex4-1}
\usepackage{subcaption}
\usepackage{bm,graphicx,tabularx,array,booktabs,dcolumn,xcolor,microtype,multirow,amscd,amsmath,amssymb,amsfonts,physics,siunitx}
\usepackage[version=4]{mhchem}
\usepackage[utf8]{inputenc}
\usepackage[T1]{fontenc}
\usepackage{txfonts}

\usepackage[normalem]{ulem}
\definecolor{hughgreen}{RGB}{0, 128, 0}

\usepackage[
	colorlinks=true,
    citecolor=blue,
    linkcolor=blue,
    filecolor=blue,      
    urlcolor=blue,	
    breaklinks=true
	]{hyperref}
\newcommand\Tstrut{\rule{0pt}{2ex}}         
\newcommand\Bstrut{\rule[-0.9ex]{0pt}{0pt}}   

\usepackage{caption} 
\usepackage{subcaption}
\usepackage{cleveref}

\usepackage{float} 
\newcommand{\Wfn}{\Psi}
\newcommand{\WfnHF}{\Wfn_\text{HF}}

\newcommand{\Zc}{Z_\text{c}}

\newcommand{\Zcfp}{Z_{\text{c}}^{\text{UHF}}}

\newcommand{\e}{\text{e}}
\newcommand{\Eh}{\text{E}_{\text{h}}}

\newcommand{\br}{\bm{r}}
\newcommand{\bx}{\bm{x}}

\newcommand{\bnabla}{\boldsymbol{\nabla}}

\renewcommand{\d}{\mathrm{d}}

\renewcommand{\H}{\mathcal{H}}
\newcommand{\hf}{\Hat{f}}
\newcommand{\hh}{\Hat{h}}
\newcommand{\hJ}{\Hat{J}}
\newcommand{\hK}{\Hat{K}}
 
\newcommand{\Nb}{n}

\usepackage{mathtools}
\newcommand{\etal}{\textit{et al.}} 

\renewcommand{\d}{\mathrm{d}}

\newcommand{\UOX}{Physical and Theoretical Chemistry Laboratory, Department of Chemistry, University of Oxford, South Parks Road, Oxford, OX1 3QZ, U.K.}

\newcommand{\SuppI}{Supporting Information}

\begin{document}

\title{Hartree--Fock Critical Nuclear Charge in Two-Electron Atoms}

\author{Hugh~G.~A.~Burton}
\email{hugh.burton@chem.ox.ac.uk}
\affiliation{\UOX}
\date{\today}

\begin{abstract}
Electron correlation effects play a key role in stabilising two-electron atoms near the critical nuclear charge, 
representing the smallest charge required to bind two electrons.
However, deciphering the importance of these effects relies on fully understanding the uncorrelated Hartree--Fock description. 
Here, we investigate the properties of the ground state wave function in the  small nuclear charge limit using various
symmetry-restricted Hartree--Fock formalisms. 
We identify the nuclear charge where spin-symmetry breaking occurs to give an 
unrestricted wave function that predicts the ionisation of an electron. 
We also discover another critical nuclear charge where the closed-shell electron density detaches from the nucleus, and 
identify the importance of fractional spin errors and static correlation in this limit.
\end{abstract}

\maketitle

\raggedbottom

\section{Introduction}
How much positive charge is required to bind two electrons to a nucleus?
This  simple question has been subject to intense research and debate ever 
since the early 1930s. %
\cite{Hylleraas1930,Stillinger1966,Stillinger1974,Ivanov1995a,Baker1990,Armour2005,Estienne2014,King2015,OlivaresPilon2015}
High-precision calculations have only recently converged on a critical nuclear charge 
for binding two electrons of 
$\Zc~=~0.911\,028\,224\,077\,255\,73(4)$.\cite{Estienne2014,King2015,OlivaresPilon2015} 
For $Z > \Zc$, the two-electron atom ($Z\, \e\, \e$) is bound and stable, with an energy lower than 
the ionised system ($Z\, \e + \e$).
However, for $Z < \Zc$, the energy of the bound atom becomes higher than 
the ionised system, causing an electron to spontaneously detach from the nucleus.
As a result, the critical charge marks the threshold for stability in the three-body problem\cite{Kais2000,Armour2005,King2013} 
and can be interpreted as a quantum phase transition.\cite{Kais2006}

While the critical nuclear charge is fascinating in its own right, the two-electron atom
also provides an essential model for understanding the performance of electronic structure approximations.
It is the simplest chemical system where electron correlation is present, 
which is thought to be essential in binding the two electrons near
$\Zc$.\cite{Stillinger1974,King2018,Baskerville2019}
In particular, comparing the closed-shell HF energy to the exact energy of the ionised system
shows that HF theory fails to predict a stable two-electron atom with $\Zc < Z < 1.031\,177\,528$,\cite{King2018}
including \ce{H-}.\cite{Goddard1968,Cox2020}
However, interpreting exactly \emph{how}
correlation influences the stability of the two-electron atom is made difficult by an incomplete
understanding of the HF approximation for small $Z$.

In many ways, placing artificial restrictions on the wave function makes  the 
HF description more complicated to interpret than the exact result.
For example, the restricted HF (RHF)  formalism can only predict
doubly-occupied orbitals,\cite{SzaboBook} and one might ask if comparing the RHF energy 
to the exact one-electron energy is a fair way to identify the RHF critical charge.
Alternatively, the unrestricted HF (UHF) approach allows the spin-up and spin-down
electrons to occupy different spatial orbitals,\cite{SzaboBook} providing a qualitatively correct 
model for the dissociation of a single bound electron in \ce{H-} at the expense of broken spin-symmetry.\cite{Cox2020,Goddard1968}
The onset of HF symmetry breaking is marked by instability thresholds in the orbital Hessian,\cite{Seeger1977}
which have also been interpreted as critical charges in closed-shell
atoms.\cite{Uhlirova2020}
These sudden qualitative changes in the HF wave function
can also be probed using the average radial electronic positions, providing an alternative
indicator for electron ionisation that does not rely on energetic comparisons with the 
exact result.
However, to the best of our knowledge, the exact nuclear charge for UHF 
symmetry breaking, and the qualitative properties of HF wave functions near this point,
remain unknown.

Previous studies on the two-electron atom using HF theory have primarily focussed on 
the large $Z$, or ``high-density'', limit (see e.g., Ref.~\onlinecite{Loos2010}).
In this limit, the closed-shell RHF wave function provides a good approximation to the exact
result, creating a model for understanding dynamic correlation.\cite{Loos2009,Loos2010}
Alternatively, the small-$Z$  ``low-density'' limit, where static correlation
becomes significant, remains far less explored. 
The primary challenges of small $Z$ include the presence of HF symmetry-breaking 
and convergence issues that occur with diffuse basis functions.
One recent HF study has been unable to reliably converge the RHF approximation for $Z < 0.85$,\cite{King2018}
hindering attempts to understand HF theory for smaller $Z$. %
Consequently, the small-$Z$ limit also provides a  model for 
understanding how to predict strong static correlation,\cite{Hollett2011} 
which remains a major computational challenge.

In this contribution, we 
investigate the properties of the RHF and
UHF ground-state wave functions in the small $Z$ limit.
We use numerical Laguerre-based HF calculations to compute the exact location of the UHF
symmetry-breaking threshold.
By investigating the average radial positions in the RHF and UHF wave functions, we assess how
each HF formalism predicts electron detachment near the critical charge.
Our results suggest that the UHF symmetry-breaking threshold represents the onset of ionisation
and forms a branch point singularity in the complex $Z$ plane.
Alternatively, RHF theory predicts a closed-shell critical point 
where half the electron density becomes ionised, leading to strong static correlation
for small $Z$.

\section{Two-Electron Atomic Hamiltonian}
%
The $Z$-scaled Hamiltonian for the two-electron atom with an infinite nuclear mass is\cite{Hylleraas1930}
\begin{equation}
\H = -\frac{1}{2} \qty(\bnabla_1^2 + \bnabla_2^2) - \frac{1}{\rho_1} - \frac{1}{\rho_2} + \frac{1}{Z} \frac{1}{\rho_{12}},
\label{eq:scaledH}
\end{equation}
where $\rho_i = r_i / Z$ is the scaled distance of electron $i$ from the nucleus,
$\rho_{12} = | \bm{\rho}_1 - \bm{\rho}_2 |$ is the scaled inter-electronic distance, 
and the unscaled distances have atomic units $a_0$.
Nuclear charges are given in atomic units $e$.
The exact wave function is defined by the time-independent Schr\"odinger equation
\begin{equation}
\H \Psi(\bx_1, \bx_2) = \tilde{E} \Psi(\bx_1, \bx_2)
\end{equation}
with the spin-spatial coordinate $\bx_i = (\bm{\rho}_i, \sigma_i)$ and the scaled energy $\tilde{E} = E / Z^2$.
The electron-electron repulsion can be considered
as a perturbation to the independent-particle model with the coupling strength $\lambda = 1/Z$,\cite{Hylleraas1930}
giving the power series expansions $\tilde{E}(\lambda) = \sum_{k=0}^{\infty} \tilde{E}^{(k)} \lambda^{k}$ and
$\Psi(\lambda) = \sum_{k=0}^{\infty} \Psi^{(k)} \lambda^k$.
The critical nuclear charge $\Zc$ can then be identified from the radius of convergence of these
series,\cite{Stillinger1966,Baker1990,Ivanov1995a,Ivanov1995b} 
defined by the distance of the closest singularity to the origin in the complex $\lambda$ plane.\cite{Marie2021}
Both $E(\lambda)$ and $| \Psi(\lambda) |^2$ have complicated singularities on the positive real axis
at $\lambda_\text{c} = 1/\Zc$,\cite{Baker1990} which have been interpreted as a quantum phase transition
in the complete-basis-set limit.\cite{Kais2006}

The HF wave function is a single Slater determinant 
$\WfnHF(\bx_1, \bx_2)$ built from the antisymmetrised product of the occupied spin-orbitals
$\psi_i(\bx)$.
These orbitals are self-consistent eigenfunctions of the one-electron Fock
operator $\hf(\bx)$, with the corresponding eigenvalues defining orbital energies.
The $Z$-scaled Fock operator is
\begin{equation}
\hf(\bx) = \hh(\bx) + \frac{1}{Z} \sum_{i=1}^{2} \qty[\hJ_i(\bx) - \hK_i(\bx)],
\label{eq:z_scaled_f}
\end{equation}
with the one-electron Hamiltonian
\begin{equation}
\hh(\bx) = -\frac{\bnabla^2}{2} - \frac{1}{\rho},
\end{equation}
and the Coulomb and exchange operators denoted 
as $\hat{J}_{i}(\bx)$ and $\hat{K}_{i}(\bx)$ respectively (see Ref.~\onlinecite{SzaboBook}).
The total HF energy is 
\begin{equation}
\tilde{E}^{\text{HF}} = \frac{1}{2} \sum_{i=1}^{2} (h_i + f_i),
\end{equation}
with the matrix elements $h_i = \langle \psi_i | \hh | \psi_i \rangle$ 
and $f_i = \langle \psi_i | \hf | \psi_i \rangle$.

The self-consistent two-electron component of the Fock operator 
can be considered as a perturbation with the coupling strength $\lambda = 1/Z$.
For large $Z$ ($\lambda \to 0$), only the one-electron component
remains and the HF wave function is exact.\cite{Loos2010} 
As $Z$ becomes smaller and $\lambda$ grows, the self-consistent repulsion
becomes increasingly dominant over the one-electron nuclear attraction.
Eventually, it becomes energetically favourable for a pair of lower-energy UHF solutions
to emerge where either the spin-up or spin-down electron becomes detached from the nucleus.\cite{Goddard1968,Cox2020}
This phenomenon is analogous to the Coulson--Fischer point in stretched \ce{H2}, where the
spin-up and spin-down electrons localise on opposite atoms,\cite{Coulson1949}
and is closely related to Wigner crystallisation.\cite{Wigner1934}
By analytically continuing an equivalent two-electron coupling parameter to complex values, we have 
recently shown that the UHF wave functions form a non-Hermitian square-root branch point at the 
symmetry-breaking threshold.\cite{Burton2019a,Marie2021}
Remarkably, following a complex-valued contour around this point leads to the interconversion of 
the degenerate solutions, and allows a ground-state wave function to be smoothly evolved into an 
excited-state wave function.\cite{Burton2019a}

\section{Computational Details}
%
In the present work, we follow Ref.~\onlinecite{King2018} and express the spatial 
component $\phi_p(\br)$ of the HF spin-orbital 
$\psi_p(\bx)$ using the spherically-symmetric Laguerre-based functions\cite{RileyBook} 
\begin{equation}
\chi_{\mu}(\br) = \exp(-\frac{A r}{2})\ L^{(1)}_{\mu}(A r),
\end{equation}
giving 
\begin{equation}
\phi_p(\br) = \sum_{\mu=0}^{\infty} \chi_\mu \qty(\br) C^{\mu \cdot}_{\cdot p}.
\end{equation}
Here we employ the nonorthogonal tensor notation of Head--Gordon \etal{}\cite{HeadGordon1998}
The non-linear parameter $A$ controls the spatial extent of the basis functions and is 
optimised alongside the coefficients $C^{\mu \cdot}_{\cdot p}$.\cite{King2018}
In practice, this expansion is truncated at a finite basis set of size $\Nb$.
To avoid previous issues with iterative solutions to the HF equations for small $Z$, 
we optimise the $C^{\mu \cdot}_{\cdot p}$ coefficients for a fixed $A$ value
using the quasi-Newton Geometric Direct Minimisation (GDM) algorithm.\cite{Voorhis2002}
The optimal $A$ value is then identified through another quasi-Newton minimisation with the 
orbital coefficients re-optimised on each step. 
All calculations were performed in a developmental version of \textsc{Q-Chem},\cite{QChem} and analytic 
expressions for the Laguerre-based integrals are provided in the \SuppI.


\section{Results}
\subsection{Spin-Symmetry Breaking Critical Point}
%
First, we identify the critical nuclear charge for HF symmetry-breaking $\Zcfp$ using a bisection
method to locate the point where the lowest orbital Hessian eigenvalue of the 
RHF solution vanishes.\cite{Seeger1977}
The convergence of $\Zcfp$ with respect to the basis set size is shown in
Table~\ref{tab:CFPconverge}, giving a best estimate of $\Zcfp = 1.057\,660\,253\,46(1)$.
This value is converged for $n \ge 24$, for which converged RHF
energies for \ce{He} and \ce{H-} are also obtained as
\begin{align*}
E_{\text{RHF}}(\text{He}) &= -2.861\,679\,995\,612\,23(1)
\\
E_{\text{RHF}}(\text{\ce{H^{-}}}) &= -0.487\,929\,734\,370\,84(1),
\end{align*} 
in agreement with the  best variational benchmarks up to 10 decimal places.\cite{King2018,Lehtola2019,Roothaan1979}
We believe that this is the first numerically precise estimate of a symmetry-breaking threshold
in the complete-basis-set HF limit, and therefore defines a new type of benchmark
value within electronic structure theory.
As expected, we find $\Zcfp > 1$, and thus our result is consistent with previous observations
of UHF symmetry breaking in the hydride anion.\cite{Cox2020,Goddard1968}

\begin{table}[thb]
\caption{Convergence of the UHF symmetry-breaking threshold $\Zcfp$ and the 
associated energy $E^{\text{UHF}}(\Zcfp)$ with respect to basis set size.
Best estimates and quoted errors correspond to the mean and standard deviation of 
the converged values $n \ge 24$ respectively.}
\label{tab:CFPconverge}
    \begin{tabular}{cS[table-format=3.16]S[table-format=2.14]}
\hline\hline
\Tstrut\Bstrut%
$\Nb$  & {$\Zcfp / e$}       & {$E^{\text{UHF}}(\Zcfp) / \Eh$}
\\ \hline
10   &    1.057 651 800 057  &  -0.570 335 516 87   \\  
12   &    1.057 658 412 462  &  -0.570 345 373 24   \\
14   &    1.057 659 966 054  &  -0.570 347 687 12   \\
16   &    1.057 660 213 291  &  -0.570 348 055 22   \\
18   &    1.057 660 248 206  &  -0.570 348 107 19   \\
20   &    1.057 660 252 818  &  -0.570 348 114 05   \\
22   &    1.057 660 253 391  &  -0.570 348 114 91   \\
24   &    1.057 660 253 461  &  -0.570 348 115 01   \\
26   &    1.057 660 253 464  &  -0.570 348 115 01   \\
28   &    1.057 660 253 462  &  -0.570 348 115 01   \\
30   &    1.057 660 253 464  &  -0.570 348 115 02   \\
32   &    1.057 660 253 439  &  -0.570 348 114 98   \\
34   &    1.057 660 253 473  &  -0.570 348 115 03   \\
36   &    1.057 660 253 458  &  -0.570 348 115 01   \\
38   &    1.057 660 253 477  &  -0.570 348 115 03   \\
40   &    1.057 660 253 466  &  -0.570 348 115 02   \\
\hline
Best &    1.057 660 253 46(1)&  -0.570 348 115 01(2) 
\\
\hline\hline
\end{tabular}
\end{table}

\begin{figure}[tbh]
\includegraphics[width=\linewidth]{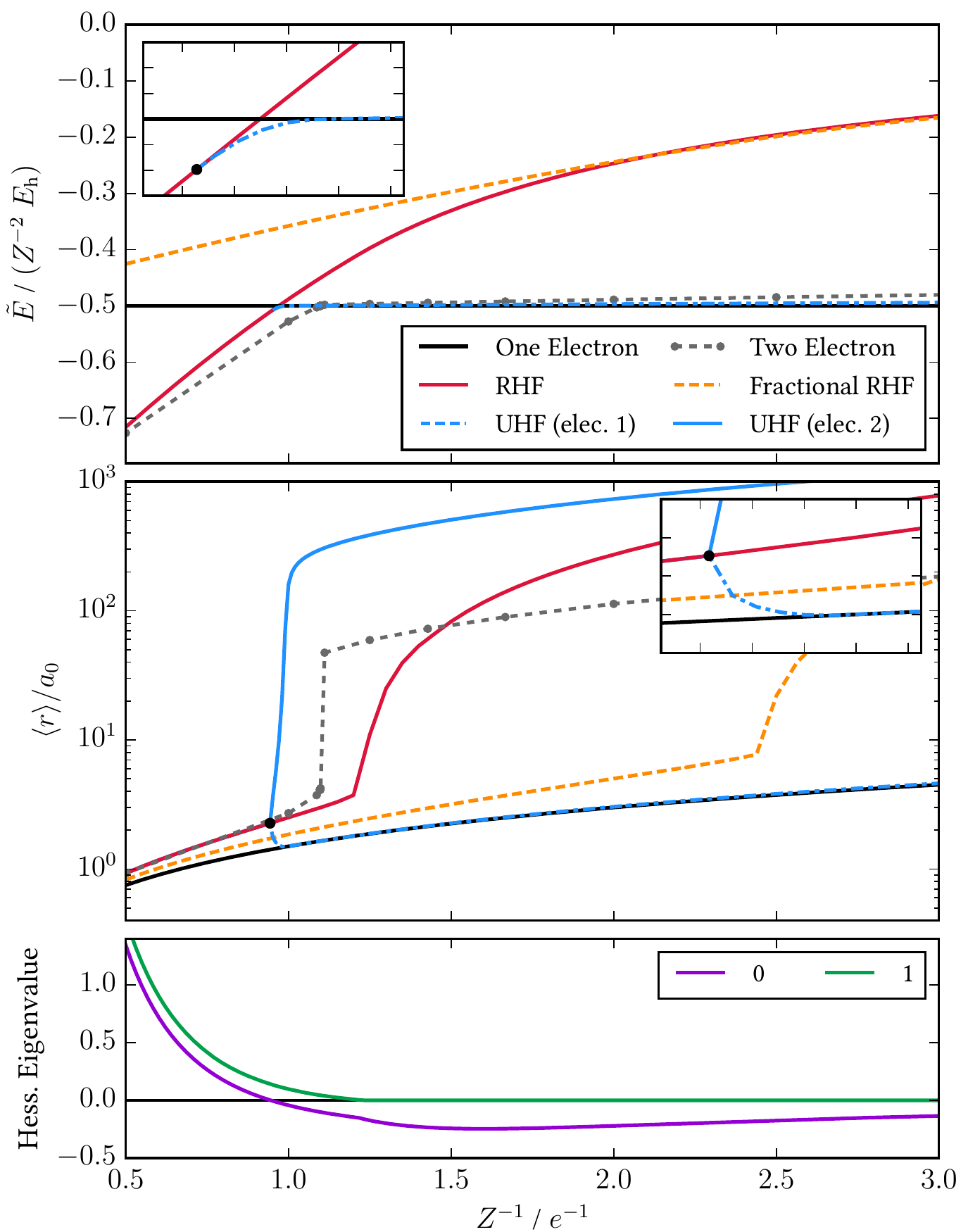}
\caption{%
$Z$-scaled energy (top) and average radial position $\langle r \rangle$ 
(middle) using various HF formalisms
and exact one- and two-electron results. 
Exact two-electron data are reproduced from Ref.~\onlinecite{King2015}.
The lowest two Hessian eigenvalues for the RHF solution (bottom) show the onset
of UHF symmetry breaking and a persistent zero eigenvalue for small $Z$.
}
\label{fig:bas26}
\end{figure}

Figure~\ref{fig:bas26} (top panel) compares the $Z$-scaled 
RHF energy (red) and the symmetry-broken UHF energy (blue dashed) as functions of $Z^{-1}$ with $n=26$.
We also consider the exact one-electron energy (black) that corresponds to the 
ionised atom, the exact two-electron energy (grey dashed; reproduced from Ref.~\onlinecite{King2015}),
and a fractional spin RHF calculation (dashed orange) with half a spin-up and half 
spin-down electron (see Sec.~\ref{subsec:FracSpinErr}). 
The UHF symmetry-breaking threshold $\Zcfp$ (black dot) occurs below the exact one-electron energy,
and thus $\Zcfp$ is greater than the HF critical nuclear charge 
previously identified using energetic arguments.\cite{King2018}
This suggests that the RHF approximation is already an inadequate representation of the exact wave function
before it becomes degenerate with the one-electron atom.
Beyond this point, the RHF energy continues to increase, while the UHF energy rapidly flattens
towards the exact one-electron result.
There is therefore a small relaxation region during which the UHF approximation
approaches a qualitative representation of the one-electron atom.

\begin{figure}[thb]
\includegraphics[width=\linewidth]{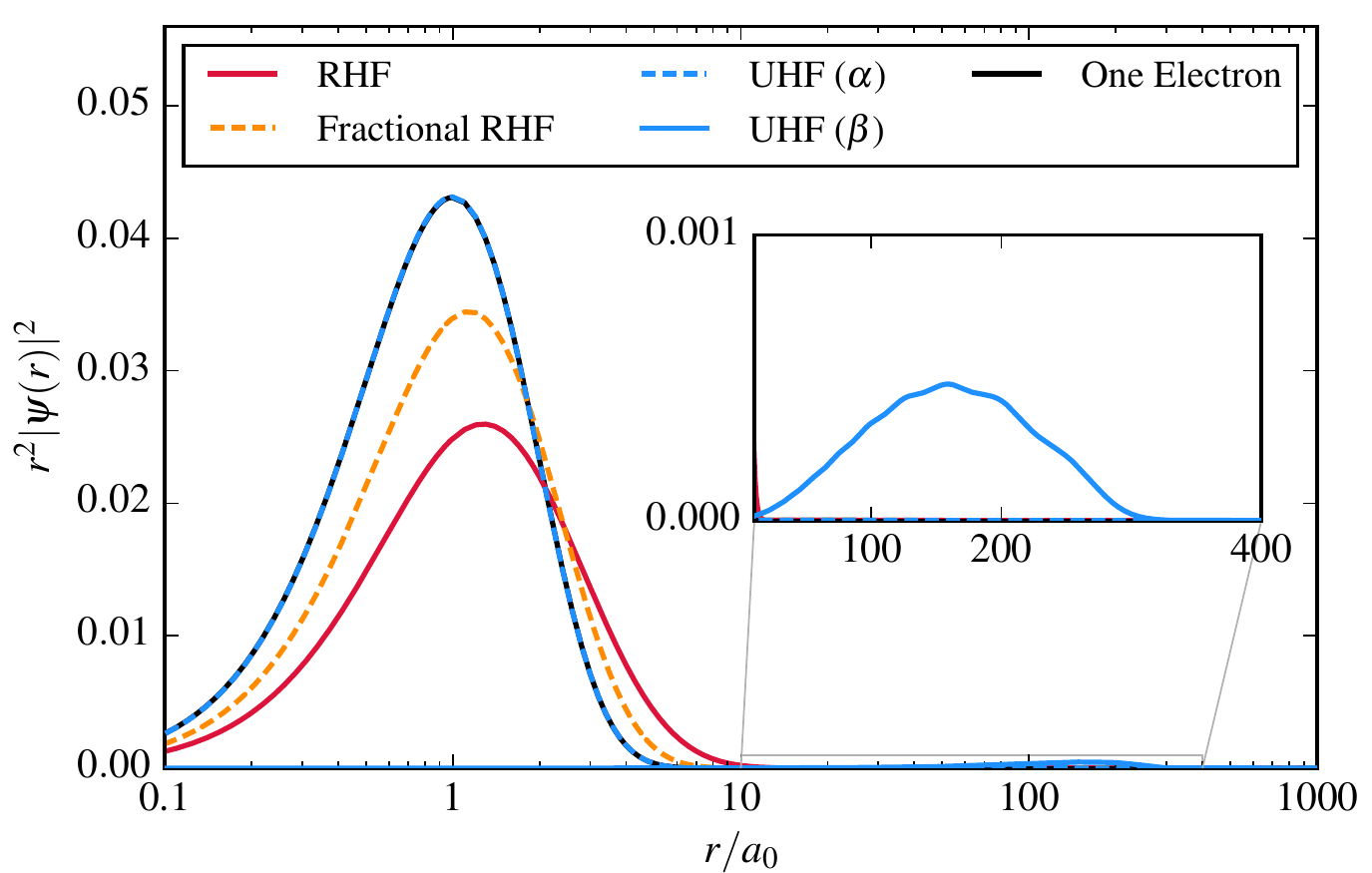}
\caption{%
Radial distribution functions for different HF orbitals compared to the exact one-electron
wave function at $Z = 1$.
}
\label{fig:radUHF}
\end{figure}

Radial electron position expectation values $\langle r \rangle$ provide further insights into the properties of the 
two-electron atom close to electron detachment.\cite{King2015,Baskerville2019,King2016}
The exact wave function yields an ``inner'' and ``outer'' 
electron, with repulsive interactions pushing the inner electron closer to the nucleus
than in the corresponding hydrogenic system.\cite{King2016} 
For $Z > \Zc^{\text{UHF}}$, the RHF radial electron position closely matches
the averaged exact two-electron result (grey dashed).
However, the RHF result starts to deviate from the two-electron
value as electron correlation effects become significant for $Z < \Zc^{\text{UHF}}$.
In contrast, the additional flexibility of the UHF wave function 
correctly predicts the separation of an inner and outer electron.
This ionisation occurs almost immediately for $Z < \Zcfp$, as indicated by a sudden increase in
$\langle r \rangle$ for the dissociating electron (Fig.~\ref{fig:bas26}: middle panel), while
the bound electron tends towards the exact one-electron result.
Comparing the radial distribution functions $P(r) = r^2 |\psi(r)|^2$ for each electronic orbital at $Z=1$ (Fig.~\ref{fig:radUHF}),
we find that the inner UHF orbital closely matches the one-electron \ce{H} atom, 
which is also the case for the exact wave function.\cite{King2016}
However, at $Z = 1$, the outer electron has essentially ionised from the atom in the UHF approximation,
but remains closely bound to the nucleus in the fully-correlated description.
The UHF wave function therefore rapidly approximates the ionised
atom for $Z < \Zcfp$, and $\Zcfp$ can be interpreted as a critical
charge for a stable two-electron atom.
This approximation essentially overlocalises the electron density between 
$\Zc < Z < \Zcfp$ (including \ce{H-}), as previously observed for two-electrons 
on concentric spheres,\cite{Loos2010b}
and fails to capture the correlation required to describe the
exact critical charge.

Figure~\ref{fig:bas26} also indicates that the dissociation of the outer electron for the UHF 
approximation is relatively sudden
and closely mirrors the behaviour at the exact critical nuclear charge.\cite{King2015}
Furthermore, there is small region where the average radial position of the inner electron
tends towards the one-electron result. 
It is known that the exact two-electron system exhibits a shape resonance as the nuclear charge
goes through the critical point, with the outer electron remaining at a finite distance from 
the nucleus.\cite{Estienne2014,Hoffmann1983}
One might therefore interpret the region where the UHF electron positions tend towards
the one-electron result as an approximation of this resonant stability regime.
 
For $Z < \Zc$, the exact wave function is an equal combination of two configurations where
either the spin-up or spin-down electron remains bound to the nucleus.
In contrast, the single-determinant nature of the UHF wave function means that only one of these
configurations can be represented: the UHF orbitals are ``pinned'' to one resonance form.\cite{Trail2003}
There must therefore be a wave function singularity at $\Zcfp$ where the UHF 
approximation branches into a form with either the spin-up or spin-down electron
remaining bound.
The mathematical structure of this point can be revealed by following a continuous pathway
around $\Zcfp$ in the complex $Z$ plane.
When $Z$ is analytically continued to complex values, the Fock operator becomes
non-Hermitian and we must consider the holomorphic HF approach.\cite{Hiscock2014,Burton2016,Burton2018}
In the remainder of this Section, we fix the non-linear $A$ parameter to its value at $\Zcfp$ as the non-Hermitian energy is
complex-valued and cannot be variationally optimised.

\begin{figure}[tbh]
\includegraphics[width=\linewidth]{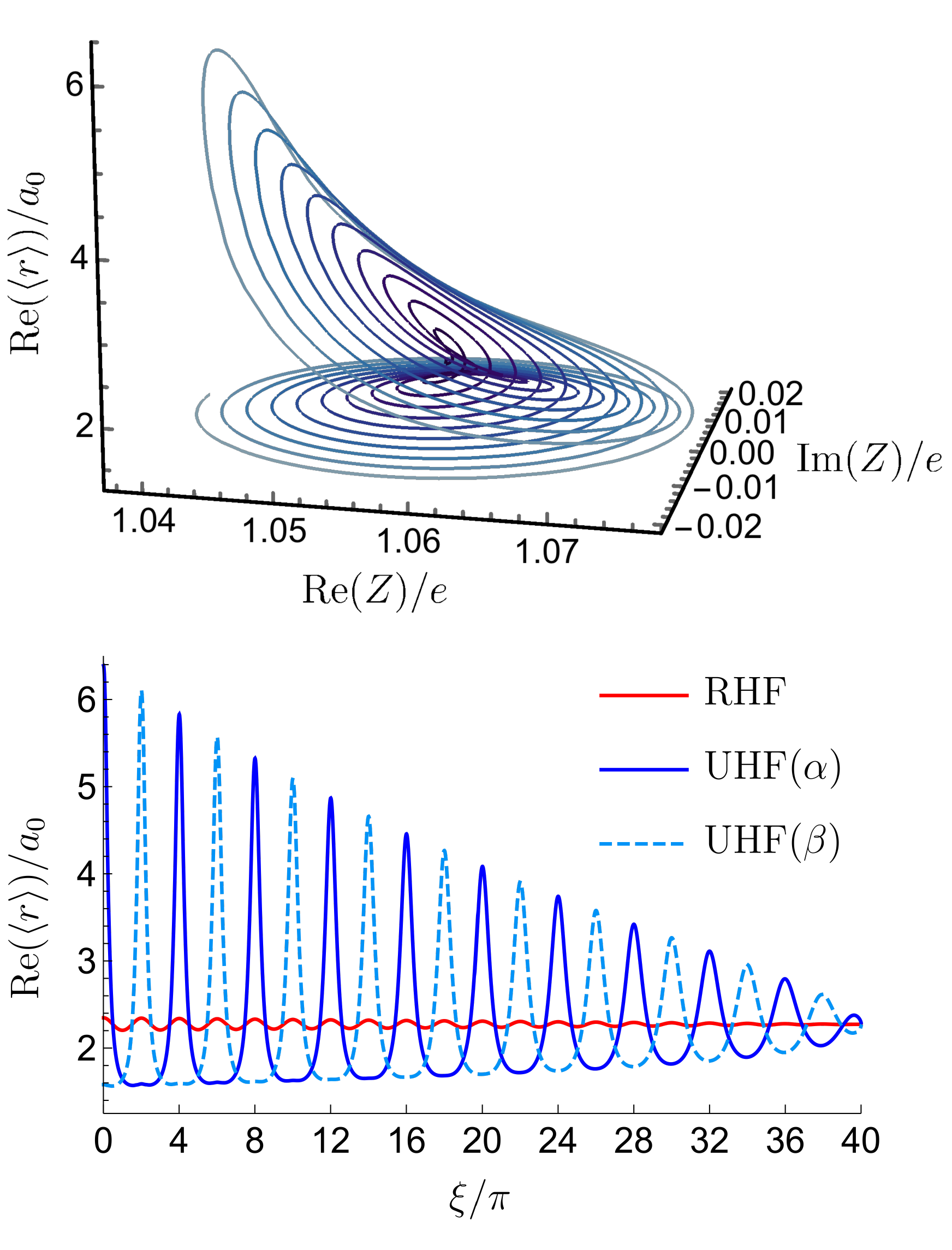}
\caption{%
Average radial position $\langle r \rangle$ of the inner electron along a spiral contour in the 
complex $Z$ plane converging on $\Zcfp$ using $\Nb = 26$.
On each rotation, the UHF wave function transitions between the two degenerate solutions.
}
\label{fig:cfp_rotate}
\end{figure}

Figure~\ref{fig:cfp_rotate} shows the real component of $\langle r \rangle$ for the (initial) inner electron 
along a pathway which spirals in towards $\Zcfp$, parametrised as
\begin{equation}
Z(\xi) = \Zcfp - \qty(0.02 - \frac{0.001 \xi}{2\pi})\exp(\text{i} \xi).
\end{equation}
Remarkably, after one complete rotation ($\xi = 2\pi$), the inner and outer electrons have swapped, 
indicating that the degenerate UHF solutions have been interconverted. 
A second full rotation is required to return the states to their original forms.
The two degenerate UHF wave functions are therefore connected
as a square-root branch point in the complex-$Z$ plane, in agreement with our previous 
observations in analytically solvable models.\cite{Burton2019a,Marie2021}
Furthermore, the branch point behaves as a quasi-exceptional point, 
where the two solutions become identical but remain normalised (see Ref.~\onlinecite{Burton2019a}),
providing the first example of this type of non-Hermitian HF degeneracy in the complete-basis-set limit.

\subsection{Closed-Shell Critical Point}
%
We now consider the fate of the RHF ground state as $Z$ continues to decrease below $\Zcfp$.
Intuitively, one might expect that the doubly-occupied RHF orbitals would 
be unable to describe the open-shell atom with an ionised electron.
Indeed, King~\etal{}\ have observed a smooth and finite $\langle r \rangle$ value for the RHF wave function
as low as $Z = 0.85$, with erratic convergence for lower nuclear charges.\cite{King2018}
A similar nuclear charge $Z=0.84$ was identified in Ref.~\onlinecite{Uhlirova2020} as a 
singlet instability threshold, where the orbital Hessian contains a zero eigenvalue with respect
to symmetry-pure orbital rotations.
These observations suggest that the RHF approximation somehow breaks down 
at $Z \approx 0.84$, but we are not aware of any detailed insight into this behaviour.

By using the gradient-based GDM algorithm,\cite{Voorhis2002} we have accurately 
converged the RHF ground state for all nuclear charges and can now firmly establish its
properties in the small-$Z$ limit.
Remarkably, we find a sudden increase in $\langle r \rangle$ at $\Zc^{\text{RHF}} = 0.82$ (Fig.~\ref{fig:bas26}: middle panel)
suggesting that the RHF approximation can, to a certain extent, represent the ionised system.
This feature closely mirrors the electron dissociation in the UHF wave function, but gives
a less sudden increase.
The smoother nature of the RHF dissociation indicates that the closed-shell restriction on the
orbitals artificially attenuates the electron detachment, providing a less accurate representation
of the exact critical charge than the UHF description.

The RHF electron detachment is accompanied by the onset of another zero eigenvalue in the orbital Hessian, 
as described in Ref.~\onlinecite{Uhlirova2020}, but we find that this 
persists for small $Z$ (Fig.~\ref{fig:bas26}: bottom panel). 
Zero Hessian eigenvalues generally indicate  a broken continuous symmetry
in the wave function, such as a global spin-rotation,\cite{Burton2021,Cui2013} and define the 
so-called ``Goldstone'' manifold of degenerate states.\cite{Cui2013,Jimenez-Hoyos2020}
In this instance, the new zero-eigenvalue Hessian mode corresponds to a spin-symmetry-breaking 
orbital rotation that also leads to an ``inner'' and ``outer'' electron.
Since the energy is constant along this mode, 
this additional zero Hessian eigenvalue suggests that the RHF approximation has 
become unstable with respect to electron detachment. 
Consequently, the sudden increase in $\langle r \rangle$ at $\Zc^{\text{RHF}}$ 
qualitatively represents a closed-shell
critical nuclear charge at $\Zc^{\text{RHF}} = 0.82$.

\begin{figure}[b]
\includegraphics[width=\linewidth]{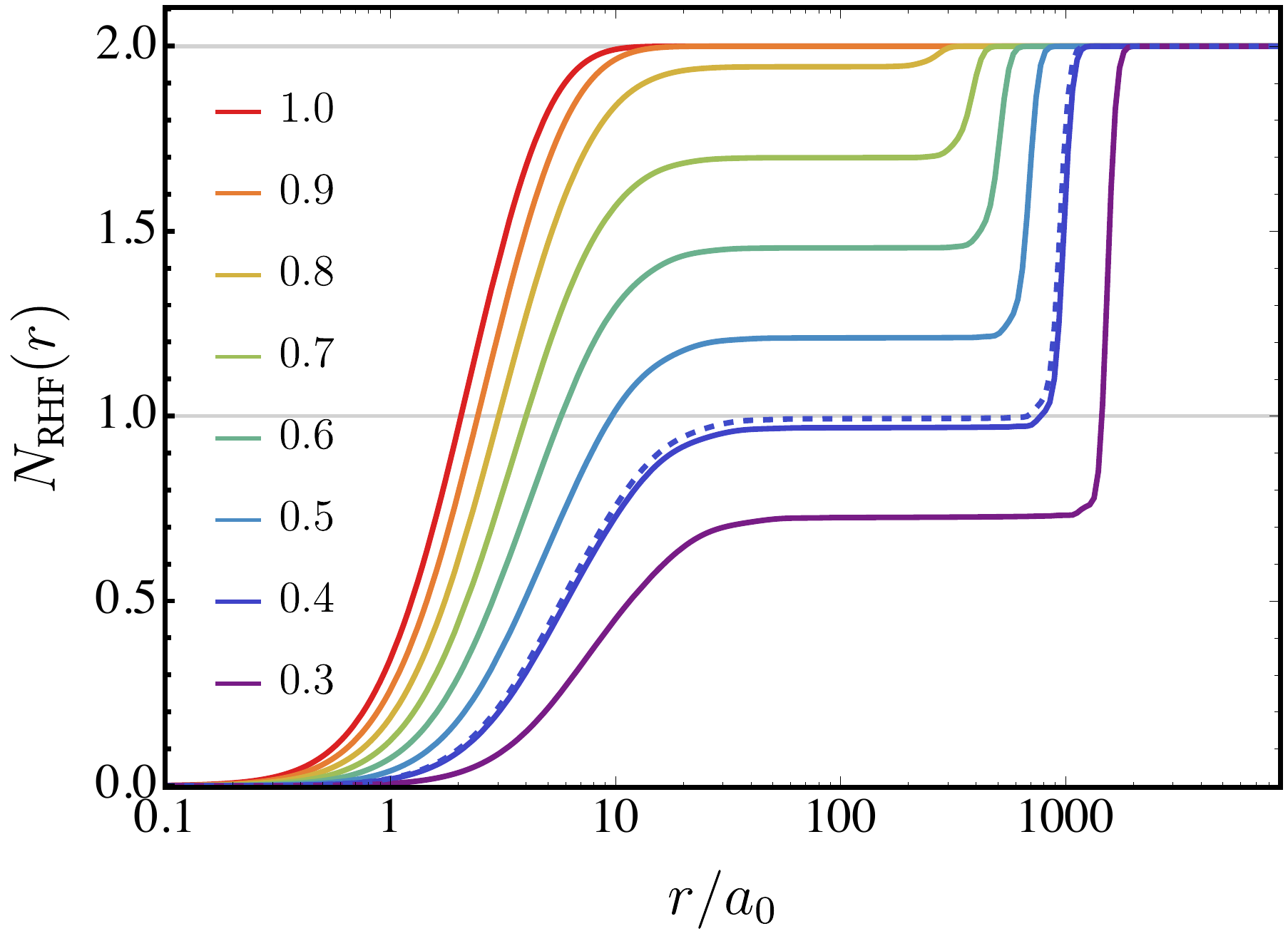}
\caption{%
Cumulative radial distribution function for the RHF wave function using $\Nb = 26$.
For $Z < 0.82$, this function adopts a double-step
structure corresponding to an inner and outer peak in the radial electron density.
At $Z = 0.41$, each peak contains half the electron density (dashed line).
}
\label{fig:rhfCDF}
\end{figure}

To further understand the electron positions in the vicinity of $\Zc^{\text{RHF}}$,
we consider the cumulative radial distribution function of the doubly-occupied RHF orbital
\begin{equation}
N_\text{RHF}(r) = \int_{0}^{2\pi} \int_{0}^{\pi} \int_{0}^{r} |\psi_\text{RHF}(r')|^2 {r'}^2 \sin \theta\, \d r'\, \d \theta\, \d \phi,
\end{equation}
as shown in Fig.~\ref{fig:rhfCDF}.
The single-step structure  at $Z>\Zc^{\text{RHF}}$ and is consistent with a single peak in the radial distribution
function (see e.g.\ Fig.~\ref{fig:radUHF}), indicating that the electrons are closely bound
to the nucleus.
For $Z < \Zc^{\text{RHF}}$, this cumulative density adopts a double-step structure 
corresponding to a radial density peak close to the nucleus, and another representing an unbound electron.
The magnitude of the second step continues to grow for smaller $Z$ as the outer peak becomes
increasingly unbound until, at $Z = 0.41$, the inner and outer peaks both contain exactly one electron.
For smaller $Z$, all the electron density becomes unbound.
Remarkably, the RHF wave function for $0.41 < Z < \Zc^{\text{RHF}}$ is therefore providing a closed-shell representation of the 
open-shell atom by delocalising the electron density over the bound and unbound radial ``sites''.
This delocalisation allows the RHF wave function to provide a qualitatively correct representation
of the exact one-body density,
but fails to capture any two-body correlation between the bound and unbound electrons.

\subsection{Fractional Spin Error}
\label{subsec:FracSpinErr}
%
Although the RHF radial density for $Z < \Zc^{\text{RHF}}$ appears to be approximating the exact result, the
RHF energy remains consistently above the one-electron hydrogenic energy. 
The closed-shell nature of the RHF orbitals means that the inner and outer radial density peaks
both contain half a spin-up electron and half a spin-down electron bound to the nucleus.
As a result, the RHF electron distribution for small $Z$ tends towards a description of the one-electron atom
that also contains half a spin-up and half a spin-down electron.
We have confirmed this limiting behaviour by computing the RHF energy with 
a half-occupied orbital, also known as the ``spin-unpolarised'' atom with 
fractional spins.\cite{Cohen2008a,Cohen2008b,Daas2020}
As expected, this half-occupied RHF solution becomes degenerate with the two-electron RHF 
energy at small $Z$ (Fig~\ref{fig:bas26}: top panel). 

Remarkably, even though a one-electron atom always has a bound ground state, 
we find that the fractional spin RHF wave function predicts an additional critical nuclear charge at 
$Z^{\text{frac}}_{\text{c}} = 0.41$,
where the (half) electrons suddenly become unbound (Fig~\ref{fig:bas26}: middle panel).
This critical charge matches the point where half the electron density has ionised
from the nucleus in the conventional RHF approach (dashed line in Fig.~\ref{fig:rhfCDF}). 
It is well-known that RHF with fractional spins fails to predict the correct energy for
one-electron atoms, despite the fact that HF theory should be exact in 
this limit, and causes the static correlation error that leads to the RHF 
breakdown for stretched \ce{H2}.
\cite{Cohen2008a,Cohen2008b}
We therefore conclude that this static correlation also creates an artificial critical 
nuclear charge in one-electron atoms at $\Zc^{\text{frac}} = 0.41$,
and is responsible for the failure of conventional RHF in the small $Z$ limit.
Identifying similar artificial critical charges using a density functional
approximation would almost certainly provide new insights into the failures of such methods for
anionic energies and electron affinities.\cite{Jensen2010,Kim2011,Peach2015}

\section{Concluding Remarks}
%
In summary, we have used average radial electronic positions to understand where HF theory predicts
electron detachment in the two-electron atom, providing alternative  critical nuclear charges in the RHF and UHF formalisms.
For UHF theory, this critical charge corresponds to a spin-symmetry-breaking 
threshold $\Zc^{\text{UHF}}~=~1.057\,660\,253\,46(1)$
where one electron suddenly ionises from the nucleus.
In contrast, at the RHF critical charge $\Zc^{\text{RHF}} = 0.82$, a secondary peak
appears in the radial distribution function at large distances from the nucleus.
%
These results provide a broader perspective on electron correlation in the small-$Z$ limit.
For example, the RHF $\langle r \rangle$ value starts to deviate from the 
exact two-electron result at the UHF symmetry-breaking threshold,
suggesting that $\Zc^\text{UHF}$ marks the onset of static correlation. 
This static correlation is further supported by the 
existence of degenerate UHF solutions representing the dominant configurations
in the exact wave function. 
Since the UHF radial distribution functions are qualitatively incorrect for 
$\Zc < Z < \Zc^\text{UHF}$, this static correlation must be essential for binding 
the two-electron atom near $\Zc$.
Furthermore, the breakdown of the half-occupied one-electron RHF result at $\Zc^\text{frac} = 0.41$
indicates that  fractional spin errors occur for small $Z$, reinforcing the importance
of static correlation.

For $Z < \Zc$, the exact wave function contains two dominant 
resonance forms with the spin-up or spin-down electron ionised from the nucleus.
As the electrons are indistinguishable, the one-electron density is delocalised
between the bound and unbound ``sites'', and this is reflected in the RHF wave function.
However, instantaneous electron-electron correlations ensure that, when one electron is bound
to the nucleus, the other electron becomes unbound. 
UHF theory provides a snapshot of these correlations, with one electron
permanently bound to the nucleus, but it cannot describe the resonance between the 
two sites. 
Alternatively, when each orbital can contain a spin-up and spin-down component in 
generalised HF (GHF), the symmetry-broken UHF solutions
form a continuum of GHF solutions parameterised by a global spin rotation.\cite{Burton2021}
The resonance between the two sites is therefore represented by this 
continuum, and could be computed using a (nonorthogonal) linear 
combination of stationary wave functions.\cite{Thom2009,Burton2019c}

Finally, the degenerate UHF wave functions form a square-root branch point 
in the complex $Z$ plane at $\Zcfp$.
Following a continuous complex path around this point interconverts the two degenerate solutions
and swaps the dissociated electron, while a second rotation returns the solutions to their original forms.
We have previously observed this behaviour in analytic models, but our current
results suggest that this phenomenon extends to the complete-basis-set limit.\cite{Burton2019a,Marie2021}
Furthermore, Ref.~\onlinecite{Burton2019a} shows that these complex branch points
can allow a ground-state wave function to be smoothly ``morphed'' into an excited-state wave function
by following a continuous complex contour.
The two-electron atom therefore provides a new model for understanding these complex connections
near the complete-basis-set limit, and we intend to continue this investigation in the future.


\section*{Supporting Information}
See the supporting information for analytic derivations of the Laguerre-based one- and
two-electron integrals.

\section*{Acknowledgements}
I gratefully thank New College, Oxford for funding through the Astor Junior Research Fellowship.
I also thank 
Hazel Cox for providing the exact two-electron numerical 
data from Ref.~\onlinecite{King2015},
and 
Pierre-Fran\c{c}ois Loos for countless inspiring conversations and critical comments on this manuscript.

\section*{Data Availability}
The data that support the findings of this study are available from the author upon reasonable request.

\bibliography{manuscript}

\end{document}